\documentclass[pra,letterpaper,twocolumn]{revtex4} 
\usepackage{amsbsy,amssymb,amsmath,bm,bbold} 
\usepackage{graphicx,color,epsfig,rotate} 
\usepackage{fancyhdr} 
\usepackage{epstopdf}
\usepackage{float}
\usepackage[colorlinks=true,linkcolor=blue,citecolor=blue,urlcolor=blue]{hyperref}

\def\bbbc{{\mathchoice {\setbox0=\hbox{$\displaystyle\rm C$}\hbox{\hbox 
to0pt{\kern0.4\wd0\vrule height0.9\ht0\hss}\box0}} 
{\setbox0=\hbox{$\textstyle\rm C$}\hbox{\hbox 
to0pt{\kern0.4\wd0\vrule height0.9\ht0\hss}\box0}} 
{\setbox0=\hbox{$\scriptstyle\rm C$}\hbox{\hbox 
to0pt{\kern0.4\wd0\vrule height0.9\ht0\hss}\box0}} 
{\setbox0=\hbox{$\scriptscriptstyle\rm C$}\hbox{\hbox 
to0pt{\kern0.4\wd0\vrule height0.9\ht0\hss}\box0}}}}

\begin{document} 
\title{Propagation and asymmetric behavior of optical pulses through time-dynamic loss-gain assisted media} 
\author{Piyali Biswas, Harsh K. Gandhi, Vaibhab Mishra,}
\affiliation{Department of Physics, Indian Institute of Technology Jodhpur, Rajasthan-342037, India}
\author{Somnath Ghosh}
\email{somiit@rediffmail.com}
\affiliation{Department of Physics, Indian Institute of Technology Jodhpur, Rajasthan-342037, India}

\begin{abstract} 
We report an asymmetric behavior of optical pulses during their propagation through a time-varying linear optical medium. The refractive index of the medium is considered to be varying with time and complex such that a sufficient amount of gain and loss is present to realize their effect on pulse propagation. We have exploited the universal formula for optical fields in time-varying media. Numerically simulated results reveal that pulses undergo opposite temporal shifts around their initial center position during their bi-directional propagation through the medium along with corresponding spectral shifts. Moreover, the peak power and accumulated chirp (time derivative of accumulated phase) of the output pulse in both propagation directions are also opposite in nature irrespective of their initial state. Numerically simulated behavior of the pulses agrees well with the analytical solutions. Possibilities have been explored in context of pulse shaping and unconventional optical devices.
\end{abstract} 
 
\maketitle %

\section{Introduction}

Asymmetric propagation of electromagnetic (EM) waves, unlike non-reciprocity, reverses the inherent properties of wave depending on the direction of propagation. Lately, such asymmetrical nature of wave transmission is of immense interest due to its wide applicability in new range of optoelectronic devices, signal processing, and to some extent in optical communication. To date, a number of investigation have been reported demonstrating asymmetric wave propagation mainly exploiting the typical characteristics of two-dimensional chiral structures, primarily metamaterials, in light-guiding systems. The first theoretical prediction and experimental observation showed that the transmission and retardation of a circularly polarized wave are different in opposite directions resulting from the planar chirality and anisotropy of a lossy medium \cite{PRL1}. This particular phenomenon were previously unknown and sooner such observations have been employed to show the asymmetrical propagation in a nanostructured metamaterial from visible to near-infrared part of the EM spectrum \cite{NanoL}. Further, a polarization independent asymmetric transmission of light has been reported by exploiting the principle of momentum symmetry breaking at discontinuous phase interfaces in a gradient index metamaterial waveguide \cite{Ncomm}. At the same time, study of asymmetric propagation has been extended to the optical fibers where the polarization rotation of EM waves are shown to be asymmetric in a spiral fiber structure that breaks the directional symmetry of the fiber and incorporates small modal losses \cite{Shemuly:13}. Very recently, towards the development of on-chip optoelectronic devices, asymmetric light transmission has been reported in hybrid plasmonic waveguide \cite{plasmon} (breaking polarization symmetry), three-layered metamaterial \cite{Tang:17} (for broad dual-band transmission in near-IR), active chiral metamaterial \cite{SciRep} (to dynamically control the wave propagation) and in quantum inspired wave based systems with exceptional singularities. However, such remarkable works have all been reported to occur in guided systems, e.g. waveguides or fibers, whereas, to the best of our knowledge, no such reports are available in literature that demonstrate asymmetric propagation of light through a bulk from implementation in device point of view. Design and fabrication of prototypes in the context of optical isolators, circulators, asymmetric mode converters are the rapidly evolving areas in integrated/ all-optical applications which directly implements such asymmetric phenomena. 

In this paper, we report an asymmetric transmission of optical pulses through a linear system by simultaneously tuning the refractive index of the medium with time. It is well known that optical pulses propagate undistorted through a linear time-invariant system. Significant temporal and spectral changes occur only when refractive index of the linear medium is time-variant \cite{yanik1,notomi,kampfrath}. Recently, both temporal and spectral shifts and pulse-shaping have been reported in time-varying refractive index media - starting from linear non-dispersive to nonlinear dispersive \cite{Xiao1,Xiao2,Xiao3,Xiao4} cases. Authors have employed a novel time-transformation technique that only considers the electric field distribution of the pulse \cite{Xiao1}. This approach is much faster than Finite difference time domain (FDTD) method, and also does not require any slowly varying envelope approximation as required in Nonlinear Schrodinger equation (NLSE) to implement other propagation techniques. Moreover, time-transformation relation can be employed to solve pulse propagation problem with any pulse duration - from short to ultrashort pulses. Here, we have numerically employed the time-transformation approach in order to study pulse propagation through the linear medium whose refractive index is time-dependent as well as complex in nature. With the imaginary part of the refractive index, we have incorporated equal and adequate amount of gain/loss such that propagating pulse will experience the effect of a gain-loss assisted media though the net gain and loss of the system is zero. We demonstrate pulse propagation in opposite directions resulting in asymmetric temporal and spectral shifts of the output pulses with quadratic phases that lead to linear chirp with opposite slopes. Moreover, power of output pulse is also affected much due to the simultaneous presence of gain and loss. Such systems once implemented will open up a new platform to develope a range of unconventional optical devices.

\section{Pulse propagation through a gain-loss assisted medium}

Propagation of optical pulses through a linear medium can be solved numerically by integrating the Maxwell's equations using the technique of finite difference time domain (FDTD) approach. However, the cumbersome nature of FDTD made researchers inquisitive about other mathematical models that provide easier approach to obtain the same output solution of the pulse propagation. Split step Fourier method has also been implemented for specific pulse shapes and durations to study the nonlinear propagation of such pulses. Recently, the time-transformation approach is reported to be the simplest mathematical model based on the relation between input and output electric field variations of the optical pulses through the equation \cite{Xiao1}
\begin{equation}
	\label{eq:eq1}
	E_{out}(t)=\int_{-\infty}^{\infty}h(t,t')E_{in}(t')dt'
\end{equation}
where, $h(t,t')$ is the impulse response function of the system. For a time-dynamic linear medium whose refractive index is varying with time, the time transformation relation is given by,
\begin{equation}
	\label{eq:eq2}
	T_r(t')=(1-s)t'/s+T_{r0}
\end{equation}
where, $s$ is the ratio of initial to final refractive index and $T_{r0}$ is the effective transit time delay. So the temporal and spectral output fields of the pulse based on the time-transformation approach are given by:
\begin{subequations}
	\begin{align}
		E_{out}(t)&=sE_{in}(st-sT_{r0}), \label{eq:subeq1} \\
		\tilde{E}_{out}(\omega)&=\tilde{E}_{in}(\omega/s)e^{-i \omega T_{r0}}. \label{eq:subeq2}
	\end{align}
\end{subequations} 
In this paper, we implement this time-transformation appropach to investigate the interesting behaviour of optical pulses through unconventional optical configurations.
\begin{figure}[htbp]
	\centering
	\includegraphics[width=\linewidth]{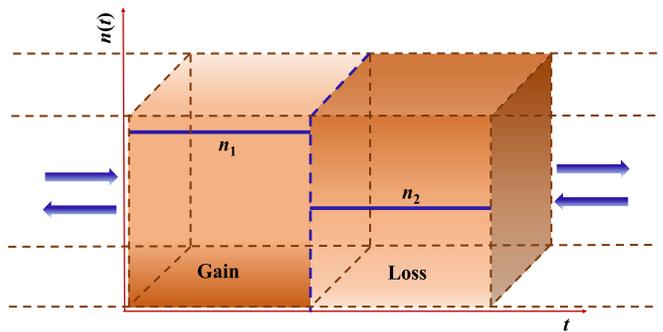}
	\caption{Schematic of the time-dynamic system. Blue arrows denote the directions of pulse propagation.}
	\label{fig:figure1}
\end{figure}
 
\subsection{Analytical approach to pulse propagation}
Here, we aim to study the propagation of optical pulses through a time-dynamic gain-loss assisted media. For this, we have considered the system to be comprised of two bulk slabs with different complex refractive indices varying temporally with a sharp index change at a specific point of time. Moreover, same amount of gain and loss has been incorporated to the system which is schematically depicted in figure \ref{fig:figure1}. The output equations (\ref{eq:subeq1}) and (\ref{eq:subeq2}) are solved analytically for an input Gaussian pulse of the form,
\begin{equation}
	\label{eq:eq3}
	E_{in}(t)=E_0e^{-t^2/2T_0^2-i \omega t}.
\end{equation}
Taking the fourier transform of Eq. (\ref{eq:eq3}), we get the frequency domain expression of the input pulse as,
\begin{equation}
	\tilde{E}_{out}(\omega)=E_0T_0\sqrt{2\pi}e^{-T_0^2(\omega-\omega_1)^2/2}.
\end{equation}
Applying the time-transformation approach we have the output form of optical pulses in both time and frequency domain:
\begin{subequations}
	\begin{align}
		E_{out}(t)&=sE_0e^{-s^2(t-T_{r0})^2/2T_0^2-is\omega_1(t-T_{r0})}, \label{eq:subeq3} \\
		\tilde{E}_{out}(\omega)&=E_0T_0\sqrt{2\pi}e^{-T_0^2(\omega-s\omega_1)^2/(2s^2)-i\omega_1T_{r0}}. \label{eq:subeq4}
	\end{align}
\end{subequations}
Now, we consider propagation of the pulse with complex refractive indices of the medium to be $n_1=n_{1r}+in_{1m}$ and $n_2=n_{2r}-in_{2m}$ respectively, which we refer to forward propagation. Hence, $s$ becomes $s=n_1/n_2=s_r+is_m$ and Eq. (\ref{eq:subeq3}) is modified to,
\begin{align}
	E_{out1}(t)=(s_r+is_m)E_0e^{(t-T_{r01})[\alpha_1(t-T_{r01})+\beta_1]} \\ \nonumber
	\times e^{-i[\alpha'_1(t-T_{r01})+\beta'_1]}, \label{eq:eq5}
\end{align}
where $\alpha_1=-(s_r^2-s_m^2)/2T_0^2$,  $\beta_1=s_m\omega_1$,  $\alpha'_1=s_rs_m/T_0^2$  and  $\beta'_1=s_r\omega_1$.
The phase of the output pulse has come out to be -
\begin{equation}
	\phi(t)=(t-T_{r01})(\alpha'(t-T_{r01})+\beta'),
	\label{eq:eq.8}
\end{equation}
from where we can find a linear dependence of instantaneous frequency on time given by,
\begin{equation}
	\delta\omega(t)=2\alpha'(t-T_{r01})+\beta'.
	\label{eq:eq9}
\end{equation}

Further, we consider backward propagation of optical pulse where the successive refractive indices of the medium are, $n_1=n_{1r}-in_{1m}$ and $n_2=n_{2r}+in_{2m}$ and hence, $s=s_r-is_m$. Accordingly the output equations becomes,
\begin{align}
	E_{out2}(t)=(s_r -is_m)E_0e^{(t-T_{r02})[(\alpha_2(t-T_{r02})+\beta_2)} \\ \nonumber
	\times e^{-i(\alpha'_2(t-T_{r02})+\beta'_2)]}
\end{align}
where, $\alpha_1=\alpha_2$, $\beta_1=-\beta_2$, $\alpha'_1=\alpha'_2$ and $\beta'_1=-\beta'_2$. 

Similarly, we can obtain the frequency domain expression for both forward and backward propagation of the output pulse:
\begin{align}
	\tilde{E}_{out1}(\omega)=E_0T_0\sqrt{2\pi}e^{-T_0^2(k\omega^2+\omega_1^2-l\omega\omega_1)/2} \\ \nonumber
	\times e^{-i(\omega_1T_{r01}+\omega(p\omega_1+q\omega)T_0^2)}, \\
	\tilde{E}_{out2}(\omega)=E_0T_0\sqrt{2\pi}e^{-T_0^2(k\omega^2+\omega_1^2-l\omega\omega_1)/2} \\ \nonumber
	\times e^{-i(\omega_1T_{r02}+\omega(p\omega_1-q\omega)T_0^2)},
\end{align}
with $k=(s_r^2-s_m^2)/(s_r^2+s_m^2)^2$, $l=2s_r/(s_r^2+s_m^2)$, $p=s_m/(s_r^2+s_m^2)$ and $q=s_rs_m/(s_r^2+s_m^2)^2$.

\subsection{Numerical observations}
Since the time-transformation approach is applicable to all kind of pulse shape and duration, thus any standard optical pulse shape from commercially available laser sources can be chosen for the study of pulse propagation. Here we have studied the evolution of a Gaussian pulse during its propagation through the time-dynamic system in opposite/both directions. For numerical simulation, we have chosen the refractive indices $n_1$ and $n_2$ to be equal to the standard silica and doped silica glasses at the operating wavelength of 1.5 $\mu$m with equal amount of gain and loss $\sim 10^{-6}$ having incorporated. 
\begin{figure}[htbp]
	\centering
	\includegraphics[width=8.5cm,height=6cm]{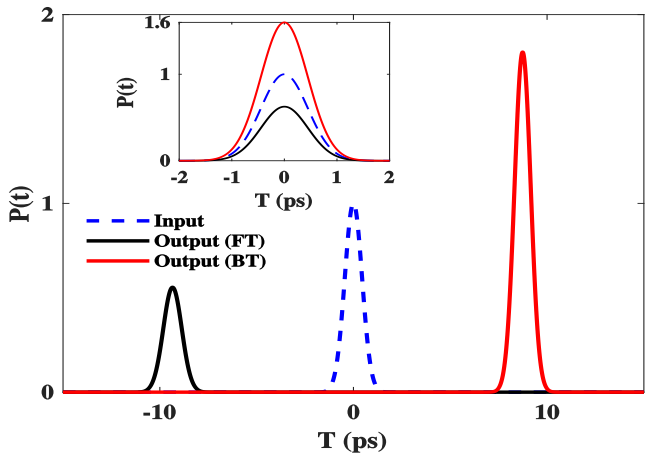}
	\caption{Time-domain input (dotted) and output pulse profile for forward (black) and backward (red) propagation in time-dynamic system. Inset shows the respective profiles for a time-static system.}
	\label{fig:figure2}
\end{figure}
The initial pulse parameters are chosen to be: peak power ($P_p$) = 100 W, full-width-at-half-maximum (FWHM) = 1 ps, and maintained the same for the bi-directional propagations. For ease of understanding of the results, we have considered the pulse propagation from left to right as forward and from right to left as backward, as depicted in figure
\ref{fig:figure1}. The temporal output profiles for both forward and backward propagations are shown in figure \ref{fig:figure2}. It is prominent that forward propagation has resulted in attenuation of the pulse whereas there is an adequate enhancement during backward propagation. Though we incorporated gain and loss in such a way that its effect will be canceled out in totality, however the real predicament is different from our notion. Such enhancement or attenuation in power occurs entirely due to the temporal shifts of output pulses. As propagated down the system, pulses have been shifted from their initial center position where $s_m$ is playing its dominant role. The respective shifts are in opposite sides of the initially fed pulse for forward ($T_{r01} = -9.3$ ps) and backward ($T_{r02} = 8.7$ ps) transmission respectively. Moreover, reshaping of pulses are evident to happen because of the complex ratio $s$. The real part of the complex ratio $s_r$ is either greater or less than $1$ depending on the respective real values of the refractive indices, and it has already been shown that pulse shape changes according to $s > 1$ or $s < 1$. As stated before, the imaginary part of $s$ i.e., $s_m$ has the dominant role to play for any possible change in pulse shape. As the output pulse has shifted from $t = 0$ to $t = T_{r0}$, thus there is a large value from $s_m$ that is contributing to the exponent part of Eq. (\ref{eq:eq5}), which is positive for forward
\begin{figure}[htbp]
	\centering
	\includegraphics[width=8.5cm,height=6cm]{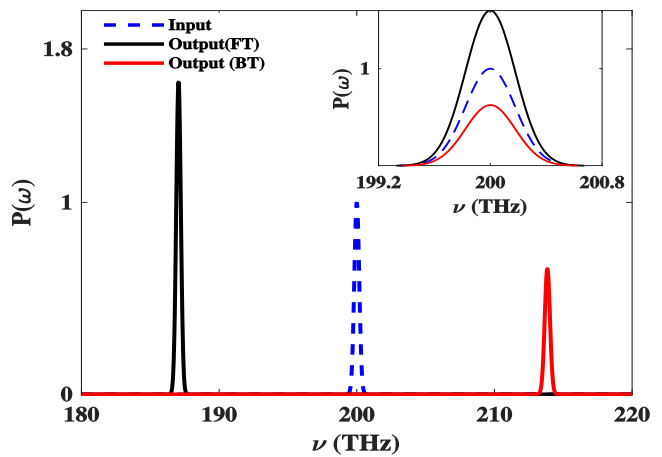}
	\caption{Input (dotted) and output spectra for both forward (black) and backward (red) propagation in time-dynamic system. Similar profiles for time-static medium are shown as inset.}
	\label{fig:figure3}
\end{figure} transmission and results in pulse attenuation, whereas negative $s_m$ during backward transmission of light enhances the power. Nearly $44\%$ of the input power has been attenuated during forward propagation through the system with its FWHM broadened by $\sim 13\%$. Whereas backward transmission has enhanced the output power by $80\%$ with $2\%$ reduction in FWHM. In this context, for a system which is time-static but gain-loss assisted, pulse propagation is asymmetric only in terms of enhancement or reduction of power as no effective time delay will be observed. This is shown as inset of figure \ref{fig:figure2}.

Accordingly, the output spectra for both propagation directions are also shifted from the input pulse centered at 200 THz with its spectral FWHM = 0.4 THz. While propagating in forward direction, output spectrum has been left-shifted by 12.9 THz from the initial center frequency, whereas 13.8 Thz shift towards right of the input frequency has been observed in opposite direction. Figure \ref{fig:figure3} clearly depicts the input and output spectra where the normalized spectral intensity has been plotted. Unlike temporal profile, forward transmission has resulted in enhancement of the spectral power by $\sim 60\%$, whereas backward propagation has attenuated the intensity by $\sim 35\%$. Moreover, no significant change has been observed in the spectral FWHM of the output pulses in opposite directions. Inset of figure \ref{fig:figure3} shows the spectral variation for time-static gain-loss assisted medium where no
\begin{figure}[htbp]
	\centering
	\includegraphics[width=8.5cm,height=6cm]{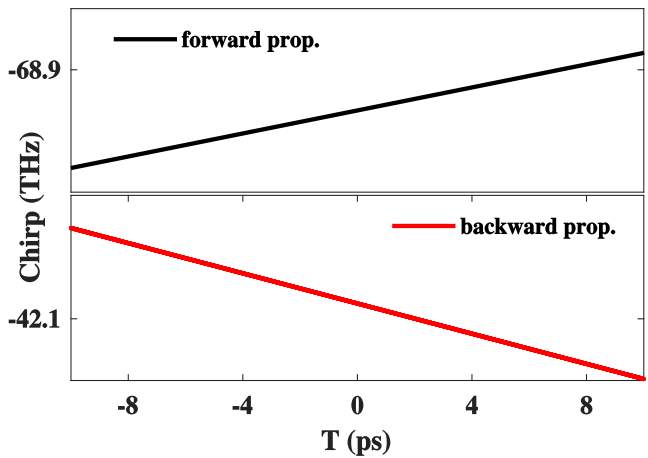}
	\caption{Variation of instantaneous frequency with time for both forward (black) and backward (red) propagation.}
	\label{fig:figure4}
\end{figure}
spectral shift has been observed. Further, according to Eq. \ref{eq:eq.8} the accumulated phase variation shows quadratic behavior which leads to a linear frequency chirp (Eq. \ref{eq:eq9}). The instantaneous frequency changes with time linearly with opposite slopes in opposite directions which corresponds to different signs of $\beta'_1$ and $\beta'_2$ and is depicted in figure \ref{fig:figure4}. 
In further to appreciate the asymmetric behaviour of the pulses, the variation of forward and backward transmission coefficients i.e., $T_f$ and $T_b$ respectively with time-shift have been estimated. In figure \ref{fig:figure5}, $logT_f$ and $logT_b$ are plotted against time-shift which clearly shows that the transmissions in opposite directions are exactly opposite in nature. $T_{r01}$ and $T_{r02}$ are the respective effective time delays where the transmission curves reach their maxima (i.e., the value of $2log|s|$ is -0.134 at $T_{r01}$ and 0.134 at $T_{r02}$). 

We further extended our investigation by increasing the number of slabs with different refractive indices. In such case, $s$ is the ratio of initial to final slab refractive indices and intermediate slabs are accounted for the computation of effective time delay $T_{r0}$. It has been observed that for a configuration with even number of slabs the propagation in opposite directions are asymmetric with changes in $T_{r0}$, whereas odd number of slabs leads to symmetric propagation with different time-shifts. Moreover, with even number of slabs the chirp is linear in both of the propagation directions, however odd numbers do not show such linearity. 
\begin{figure}[htbp]
	\centering
	\includegraphics[width=8.5cm,height=6cm]{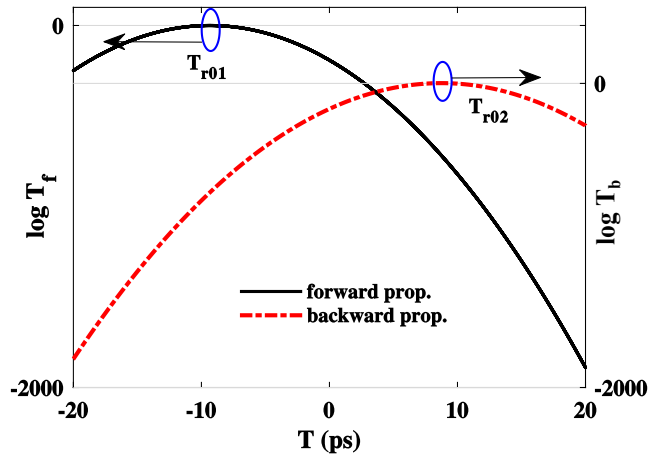}
	\caption{Logarithmic variation of forward transmission coefficient $T_f$ and backward transmission coefficient $T_b$ with related time-shift.}
	\label{fig:figure5}
\end{figure}

\section{Conclusion}

In summary, we have established an asymmetric propagation of optical pulses through a linear non-dispersive time-dynamic gain-loss assisted optical medium. To realize such unconventional behavior, we need not necessarily break any kind of symmetry or to have any synthetic material as a medium. The time-dependent refractive index with adequate amount of gain and loss in a linear medium results in asymmetric pulse propagation in opposite directions. Such an observation is very new and will eventually propel the development of unconventional optoelectronic devices, dynamic pulse shaping and spectrum tailoring.  

\section*{Acknowledgments} 

SG acknowledges support from IFA-12, PH-23 grant from DST, India and  Science and Engineering Research Board (SERB), India [ECR/2017/000491]. PB acknowledges support from MHRD.

\end{document}